\documentclass[pdflatex,sn-nature]{sn-jnl}

\usepackage{graphicx}%
\usepackage{multirow}%
\usepackage{amsmath,amssymb,amsfonts}%
\usepackage{amsthm}%
\usepackage{mathrsfs}%
\usepackage[title]{appendix}%
\usepackage{xcolor}%
\usepackage{textcomp}%
\usepackage{manyfoot}%
\usepackage{booktabs}%
\usepackage{algorithm}%
\usepackage{algorithmicx}%
\usepackage{algpseudocode}%
\usepackage{listings}%
\usepackage{todonotes}%

\theoremstyle{thmstyleone}%
\theoremstyle{thmstyletwo}%

\theoremstyle{thmstylethree}%

\usepackage{microtype}
\usepackage{hyperref}
\usepackage{subcaption}

\newcommand{\rx}{R_X}
\newcommand{\ry}{R_Y}
\newcommand{\rz}{R_Z}
\newcommand{\CRX}{\mathrm{CRX}}
\newcommand{\sampler}{\textsc{SamplerV2}}

\begin{document}

\title[Article Title]{Image Classification on IBM Quantum Computers}


\author[1]{\fnm{Junghoon Justin} \sur{Park}}\email{utopie9090@snu.ac.kr}

\author[1,2,3]{\fnm{Jiook} \sur{Cha}}\email{connectome@snu.ac.kr}

\author[4]{\fnm{Jun-gyeong} \sur{Park}}\email{0712038@yonsei.ac.kr}

\author[5]{\fnm{Hwidong} \sur{Yoo}}\email{hwi.dong.yoo@cern.ch}

\author*[6]{\fnm{Kwangmin} \sur{Yu}}\email{kyu@bnl.gov}

\affil[1]{\small \orgdiv{Interdisciplinary Program in Artificial Intelligence}, \orgname{Seoul National University}, \city{Seoul}, \country{Republic of Korea}}

\affil[2]{\small \orgdiv{Department of Psychology}, \orgname{Seoul National University}, \city{Seoul}, \country{Republic of Korea}}

\affil[3]{\small \orgdiv{Department of Brain and Cognitive Sciences}, \orgname{Seoul National University}, \city{Seoul}, \country{Republic of Korea}}

\affil[4]{\small \orgdiv{Department of Applied Statistics}, \orgname{Yonsei University}, \city{Seoul}, \country{Republic of Korea}}

\affil[5]{\small \orgdiv{Department of Physics}, \orgname{Yonsei University}, \city{Seoul}, \country{Republic of Korea}}

\affil[6]{\small \orgdiv{Computational Science Department}, \orgname{Brookhaven National Laboratory}, \state{New York}, \country{USA}}


\abstract{Quantum machine learning on real noisy intermediate-scale quantum (NISQ) hardware has remained largely confined to binary or few-class tasks, limited by the cost of on-hardware training and the underuse of large devices at inference. We present a unified framework that classifies ten-class MNIST end-to-end on a $127$-qubit IBM Eagle processor, with three central contributions. First, a two-phase protocol decouples a gradient-based classical optimization of the encoder and readout from a gradient-free optimization of the quantum parameters, removing the parameter-shift gradient cost that makes on-hardware training impractical. Second, we introduce Quantum Multi-Programming to a trained quantum classifier for the first time, packing multiple circuit copies onto one device to deliver parallel inference at no mean-accuracy cost while cutting quantum-processor job submissions proportionally. Third, a controlled comparison shows that on-hardware fine-tuning yields no measurable accuracy gain, motivating a practical NISQ workflow: train on a classical simulator and reserve the hardware for inference only. Benchmarked against a matched-capacity classical network, the quantum module shows no per-parameter accuracy advantage at this scale; we therefore frame the work as a feasibility-and-workflow demonstration for multi-class quantum image classification on current hardware.}




\maketitle

\section{Introduction}\label{sec1}

Recent years have witnessed a paradigm shift in quantum computing, with hardware and algorithmic advancements enabling successful simulations that scale well beyond simple toy systems into the one-hundred-qubit regime. Quantum platforms have increasingly demonstrated the capacity to emulate complex, classically intractable many-body phenomena, ranging from strongly correlated electron dynamics in the Fermi-Hubbard model to frustrated quantum magnetism in $J_1$-$J_2$ systems and gauge theories via the Schwinger model \cite{chowdhury2026FHM, chowdhury2024enhancing, farrell2024quantum}. Notably, real-device executions on architectures like the 127-qubit IBM Eagle processor have provided concrete evidence for quantum utility by outperforming brute-force classical simulation methods in tracking the time dynamics of non-trivial spin systems before the advent of full fault tolerance~\cite{kim2023evidence}.

Driven by this rapid progress, quantum computing has attracted growing attention as a potential accelerator for machine learning. Parameterized quantum circuits (PQCs) can define feature maps that operate in exponentially large Hilbert spaces, which are classically intractable to simulate efficiently~\cite{havlicek2019supervised}. Among the most actively studied architectures are variational quantum classifiers (VQCs), which optimize shallow parameterized circuits for classification through classical--quantum hybrid feedback loops~\cite{farhi2018classification}, and quantum convolutional neural networks (QCNNs), which translate the hierarchical pooling and convolution structure of classical CNNs into unitary quantum operations~\cite{cong2019qcnn}. Both architectures are designed to operate within the constraints of current hardware. Despite this progress, translating quantum machine learning (QML) from simulation to execution on real noisy intermediate-scale quantum (NISQ) devices remains a substantial challenge~\cite{preskill2018quantum}, with difficulties spanning classification scale, training cost, inference throughput, and architecture selection.

A central bottleneck is the limited scale of classification tasks that have been demonstrated on actual quantum hardware. The majority of hardware-validated QML experiments restrict themselves to binary classification, and the few attempts at multi-class settings reveal a consistent pattern across different datasets, architectures, and devices: accuracy degrades sharply as the number of classes increases. R\"oseler \textit{et al.}~\cite{roseler2025efficient} achieved 96.08\% accuracy on a binary MNIST task using a 49-qubit fragment-encoding QCNN on IBM's Heron r2 processor, but reported substantially lower accuracy (55.1\%) when extending to four classes. Shen \textit{et al.}~\cite{shen2024classification} deployed variational classifiers for ten-class Fashion-MNIST on IBM's ibmq-kolkata and reported hardware accuracies near 40\%, further illustrating the challenge of scaling beyond binary settings on current devices. Singh \textit{et al.}~\cite{singh2026benchmarking} benchmarked purely quantum classifiers across eight MedMNIST medical imaging datasets on IBM's 127-qubit Cleveland processor, achieving 85.4\% accuracy on a binary task (PneumoniaMNIST) and 41.8\% and 34.0\% on nine- and eleven-class problems respectively, even after applying error suppression and mitigation. These results collectively suggest that achieving reliable multi-class image classification on NISQ hardware remains an open problem.

A second challenge---one that further limits the ability to close this performance gap---is the practical cost of training quantum circuits on hardware. Gradient-based optimization of PQC parameters typically requires the parameter-shift rule, which demands $2N$ circuit executions per gradient step for $N$ trainable parameters~\cite{mitarai2018quantum, schuld2019evaluating}. On cloud-accessed quantum processors, where each circuit submission incurs significant queue latency and shot overhead, this cost renders end-to-end on-hardware training impractical for all but the smallest circuits. The problem is further aggravated by the barren plateau phenomenon: for randomly initialized circuits, cost-function gradients vanish exponentially with system size, effectively stalling optimization as the number of qubits or circuit depth grows~\cite{mcclean2018barren, cerezo2021variational}. Although several mitigation strategies have been proposed, including layerwise training~\cite{skolik2021layerwise} and locality-preserving ans\"atze~\cite{pesah2021absence}, a practical training protocol that decouples the expensive quantum gradient computation from the classical optimization loop has not been systematically explored for hardware-deployed image classifiers.

A third limitation concerns inference throughput---the stage that, given the impracticality of on-hardware training, constitutes the primary interaction between QML models and real quantum devices. Current NISQ processors, such as IBM's 127-qubit Eagle-generation devices, possess far more physical qubits than a typical QML circuit requires, yet standard practice executes one circuit per job submission. The resulting wall-clock time is dominated by per-job overhead rather than quantum computation, leaving the vast majority of the device idle. Quantum Multi-Programming (QMP)~\cite{das2019case, niu2023enabling, park2023qmp, baker2024qmp}---packing multiple independent circuit instances onto a single wide device---has been explored in the context of variational eigensolvers and combinatorial optimization, but its application to QML inference pipelines has received little attention.

A fourth challenge cuts across the preceding three: all of them require choosing an appropriate circuit architecture, yet this decision itself lacks empirical grounding. While numerous ansatz designs have been proposed for quantum classifiers, such as circular-entanglement VQCs, brick-wall topologies, and hierarchical QCNNs, comparative evaluations are typically reported under heterogeneous training conditions and rarely pair classification performance with the hardware-relevant compilation cost (circuit depth and entangling-gate count) that ultimately governs feasibility on a noisy device. The practical question of which ansatz topology offers the best accuracy--cost trade-off for hardware deployment therefore remains largely unanswered, because answering it requires a controlled comparison under a single training, data, and mitigation protocol with post-transpilation cost metrics reported on the same footing.

In this work, we address all four limitations through a unified experimental framework deployed end-to-end on IBM's 127-qubit Eagle-r3 processor (ibm\_yonsei), using ten-class MNIST handwritten-digit classification as the evaluation task. Our contributions are as follows:

\begin{enumerate}
    \item[(i)] We present a unified framework that jointly addresses training, inference efficiency, and architecture selection for multi-class quantum image classification on real hardware---an integration that, to our knowledge, has not been attempted in prior work. We evaluate the framework on the ten-class MNIST problem, going beyond prior hardware studies that tackle these challenges in isolation or restrict themselves to binary and few-class settings.

    \item[(ii)] We propose a two-phase hybrid training protocol that separates classical and quantum optimization. In Phase 1, the classical encoder and readout parameters are trained via gradient descent on a noiseless simulator while the quantum parameters remain frozen, treating the quantum circuit as a fixed nonlinear feature map. In Phase 2, only the quantum parameters are optimized using the gradient-free COBYLA algorithm, which avoids the $2N$-circuit overhead of parameter-shift gradients and is inexpensive enough to be executed directly on hardware. We then use this protocol to ask whether on-hardware fine-tuning is worthwhile: in a controlled comparison, two epochs of on-hardware COBYLA optimization yield no measurable accuracy improvement over training entirely on a simulator. This negative result directly motivates an inference-only deployment strategy---training the model classically and reserving the quantum hardware for the forward pass---which we develop in contributions (iii) and (v).

    \item[(iii)] We integrate Quantum Multi-Programming into the inference pipeline, packing $K=4$ identical 12-qubit circuit copies onto the 127-qubit device in a single submission. We prove that QMP is mathematically equivalent to serial execution in the noiseless limit, and find empirically that $K=4$ packing leaves the mean test accuracy unchanged within sampling noise---amplifying cross-seed variance rather than degrading the mean---while reducing the number of quantum processing unit (QPU) job submissions $K$-fold. Because per-sample wall time on cloud-accessed hardware is dominated by per-job queue and submission overhead rather than quantum computation, this job-count reduction yields a structural throughput gain for the inference stage.

    \item[(iv)] We conduct a controlled comparison of five variational ans\"atze---three VQC entanglement topologies and two hierarchical QCNN variants---under identical data splits, training protocols, and error-mitigation settings in noiseless simulation, pairing classification accuracy with post-transpilation circuit metrics (depth and CNOT count) computed for the IBM native gate set. The selected ansatz is then deployed on hardware, enabling practitioners to make informed architecture choices for NISQ deployment on the basis of both classification performance and compilation cost.

    \item[(v)] Synthesizing the above, we establish a concrete and practical NISQ workflow---train the quantum model on a classical simulator, then deploy it on hardware for inference only---validated by a controlled five-cell experimental design that isolates the effects of hardware noise, on-hardware training, and QMP packing while holding bit-identical trained weights fixed across the compared cells. To calibrate this result honestly, we benchmark against a matched-capacity classical multilayer perceptron and find no per-parameter accuracy advantage at this scale; we therefore frame our contribution as a feasibility-and-workflow demonstration for multi-class QML on current hardware, rather than a claim of quantum accuracy advantage.
\end{enumerate}

\section{Methods}\label{sec:methods}

\subsection{Overview and notation}

We evaluate a hybrid classical--quantum classifier on the ten-class MNIST handwritten-digit benchmark, deployed end-to-end on a current superconducting quantum processor (IBM Eagle generation, $127$ physical qubits; backend \texttt{ibm\_yonsei}).
The classifier shares a single architectural skeleton:
\begin{equation}
\mathbf{x}\in\mathbb{R}^{784}
~\xrightarrow{\text{classical encoder}}~
\mathbf{z}\in\mathbb{R}^{n_q}
~\xrightarrow{\text{VQC}~U(\boldsymbol{\theta})}~
\langle Z\rangle\in[-1,1]^{n_q}
~\xrightarrow{\text{classical readout}}~
\mathbf{y}\in\mathbb{R}^{10},
\label{eq:pipeline}
\end{equation}
where $n_q=12$ is the number of qubits, the encoder is a single \textsf{Linear}$(784\!\to\!n_q)$ map, the quantum circuit $U(\boldsymbol{\theta})$ is a parameterized ansatz that takes $\mathbf{z}$ as rotation angles via a $Y$-axis angle embedding, $\langle Z\rangle$ denotes the vector of single-qubit Pauli-$Z$ expectation values measured at the output, and the readout is a single \textsf{Linear}$(n_q\!\to\!10)$ map producing class logits.
$\boldsymbol{\theta}$ denotes the trainable \emph{quantum} parameters and $\boldsymbol{\phi}$ denotes the classical $\mathsf{Linear}$ weights.
The architecture in Eq.~\eqref{eq:pipeline} keeps the quantum subsystem small enough to execute on contemporary NISQ hardware while letting the classical $\boldsymbol{\phi}$ handle the high-dimensional input projection.
Quantum parameters comprise $0.16\%$ of the total parameter count ($15$ quantum versus $9{,}550$ classical), a capacity asymmetry that we address explicitly via a matched-capacity classical baseline (see ``Matched-capacity classical baseline'' below) and discuss in the Discussion.

\subsection{VQC\_ZigZag architecture}\label{sec:zigzag-arch}

\begin{figure}[t]\centering
\includegraphics[width=\textwidth]{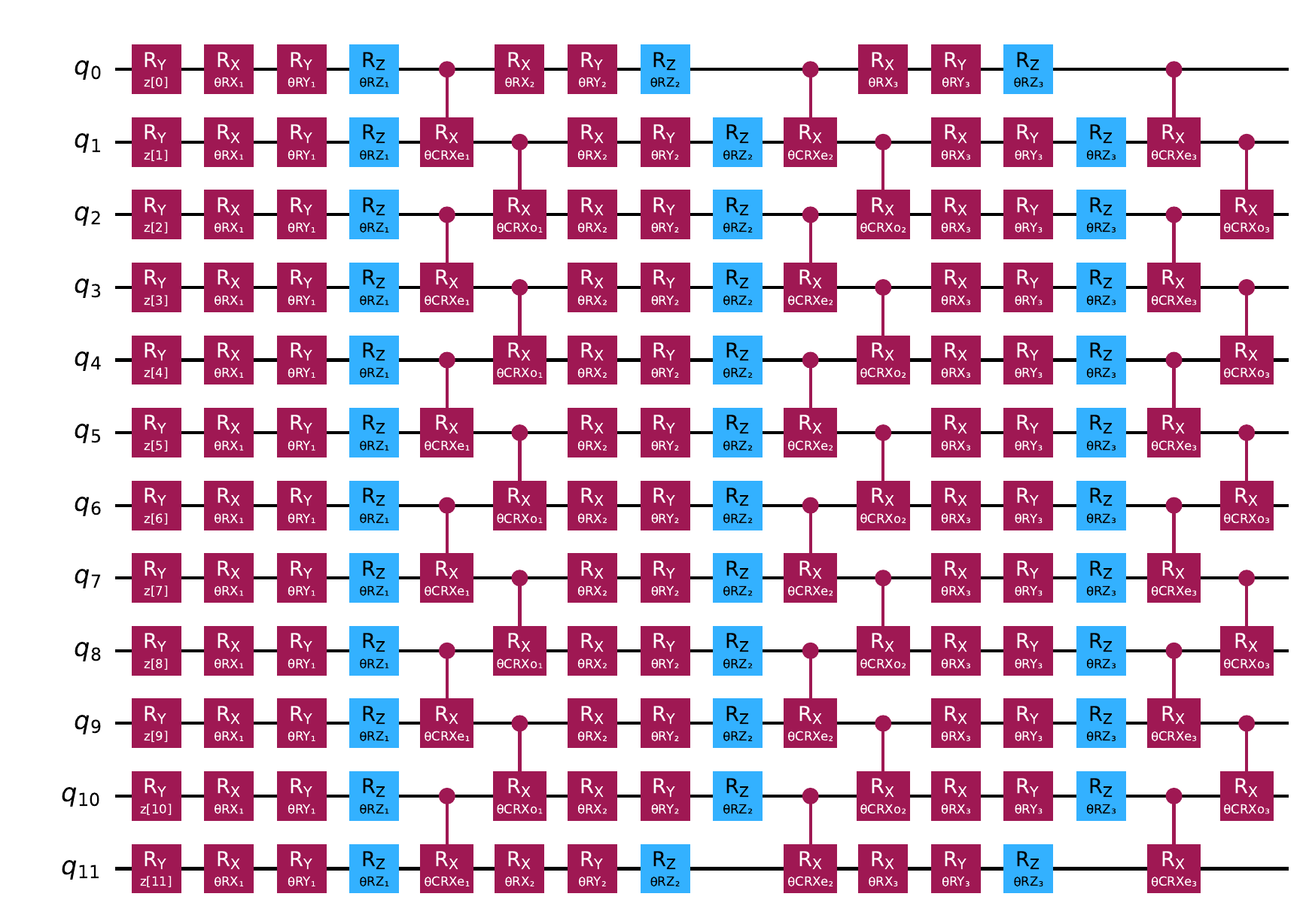}
\caption{Logical VQC\_ZigZag circuit ($n_q=12$, $L=3$): a $Y$-axis angle embedding ($\ry$) of the encoder output, then $L$ layers of single-qubit $\rx,\ry,\rz$ rotations and two brick-wall $\CRX$ sublayers (even then odd adjacent pairs, no wrap-around), and a Pauli-$Z$ readout on all $12$ qubits. Gate labels are symbolic rather than numerical: $z_i$ is the encoder output angle applied to qubit $i$ (twelve distinct, sample-dependent values), and $\theta_{\rx}^{(\ell)}$, $\theta_{\ry}^{(\ell)}$, $\theta_{\rz}^{(\ell)}$, $\theta_{\CRX,\mathrm{even}}^{(\ell)}$, $\theta_{\CRX,\mathrm{odd}}^{(\ell)}$ are the five trainable angles of layer $\ell$; the figure carries the layer index as a subscript and abbreviates the even- and odd-pair sublayers as $e$ and $o$. Each angle is shared across all twelve qubits of its layer, giving five trainable parameters per layer and $L\cdot 5 = 15$ in total.}
\label{fig:zigzag-logical}\end{figure}

VQC\_ZigZag was selected from a five-ansatz pilot comparison (Sec.~\ref{sec:pilot}) on the basis of test accuracy, cross-seed consistency, and post-transpilation cost; it is the model deployed in all hardware experiments.
VQC\_ZigZag operates on $n_q=12$ qubits with $L=3$ layers.
Per layer, the ansatz applies single-qubit $\rx(\theta_{\rx}^{(\ell)})$, $\ry(\theta_{\ry}^{(\ell)})$, $\rz(\theta_{\rz}^{(\ell)})$ rotations on every qubit, followed by two entangling sublayers of $\CRX$ gates on \emph{even} then \emph{odd} adjacent qubit pairs, with one shared parameter per sublayer ($\theta_{\CRX,\mathrm{even}}^{(\ell)}$, $\theta_{\CRX,\mathrm{odd}}^{(\ell)}$) and no wrap-around coupling between the last and first qubits.
The total trainable quantum-parameter count is $L\cdot 5 = 15$.
The post-transpilation gates-only depth on the IBM basis is $45$ with $66$ CNOTs under the heaviest available level of compiler optimization (Table~\ref{tab:pilot-metrics}); the brick-wall CRX entanglers sit on distributed, non-consecutive qubit pairs, so the CNOT count is invariant to the optimization level and only the single-qubit-gate depth is reduced by compilation.
The logical circuit is shown in Fig.~\ref{fig:zigzag-logical}.

\subsection{Two-phase hybrid training protocol}

Joint optimization of $(\boldsymbol{\phi},\boldsymbol{\theta})$ on hardware is impractical because gradients of $\boldsymbol{\theta}$ require parameter-shift evaluations ($2N$ circuit executions per gradient step for $N$ parameters), incompatible with QPU queue latency and shot cost.
Joint optimization on a noiseless simulator suffers from barren-plateau pathologies for sufficiently expressive quantum circuits.
We use a two-phase protocol that decouples the classical and quantum optimizations.

\subsubsection{Phase~1 (gradient-based, classical layers only).}
The classical parameters $\boldsymbol{\phi}$ are trained with the Adam optimizer (learning rate $10^{-3}$, weight decay $10^{-4}$) for $30$ epochs on a noiseless statevector simulator that returns analytical expectation values.
The quantum parameters $\boldsymbol{\theta}$ are frozen at random initialization ($\boldsymbol{\theta}\sim 0.1\cdot\mathcal{N}(0,1)$) and not optimized.
Phase~1 produces a checkpoint $(\boldsymbol{\phi}^\star,\boldsymbol{\theta}^{\mathrm{init}})$ in which the encoder/readout has been fitted to a random fixed quantum feature map.
Phase~1 is identical across all hardware experiments (same Phase~1 checkpoint per seed feeds every downstream Phase~2 configuration).

\subsubsection{Phase~2 (gradient-free, quantum parameters only, Sampler-everywhere).}
Beginning from the Phase~1 checkpoint, $\boldsymbol{\phi}^\star$ is frozen and only $\boldsymbol{\theta}$ is optimized.
Optimization uses the COBYLA algorithm~\cite{powell1994cobyla} via the Nevergrad gradient-free optimization framework~\cite{rapin2018nevergrad}, which avoids the gradients that would otherwise require parameter-shift evaluations.
The COBYLA budget is $B$ function evaluations over $E$ epochs, yielding $\max(1,\lfloor B/E\rfloor)$ COBYLA iterations per epoch.

Phase~2 runs in two distinct training-time device configurations that share the same measurement primitive (the Qiskit Runtime \sampler{} primitive~\cite{ibm_qiskit_runtime} with $256$ shots per circuit) and the same mitigation pipeline (dynamical decoupling~\cite{Viola1999DD,Ezzell2023DD} on idle qubits and Pauli-twirled measurements; see ``Quantum hardware deployment and error mitigation'' below).
This ``Sampler-everywhere'' design eliminates any cross-primitive mitigation-strength confound between training and inference.

\begin{itemize}
\item \emph{Aer-noiseless Phase~2}: all $E=20$ epochs execute on a noiseless Aer simulator (with hardware noise disabled), routed through the same backend interface as the on-device experiments. The only difference between this configuration and on-device execution is the absence of hardware noise; finite-shot Monte Carlo noise is present in both. The Aer simulator's shot-sampling random state is seeded explicitly so that the run is bitwise reproducible. The COBYLA budget is $B=500$ ($25$ iterations per epoch).
\item \emph{Hybrid Phase~2 (Sampler)}: epochs $1$--$18$ execute on Aer-noiseless via \sampler{}, and epochs $19$--$20$ execute on the QPU \texttt{ibm\_yonsei} via \sampler{} within a per-epoch Qiskit Runtime Session capped at $10$~h of cumulative session time. The COBYLA budget is $B=20$ (one COBYLA iteration per epoch). The reduced budget for the hardware portion is dictated by per-iteration wall-time on hardware: a single iteration evaluates the loss over $\lceil N_{\mathrm{train}}/16\rceil=22$ mini-batches, each of which incurs per-job queue and execution overhead.
\end{itemize}

In both configurations, the COBYLA optimizer's internal random state is seeded explicitly so that the search trajectory is reproducible across runs.

\subsection{Quantum Multi-Programming for K-parallel inference}\label{sec:qmp}

To maximize throughput on the IBMQ system, we utilize QMP for parallel circuit execution, which allows for the concurrent processing of multiple quantum circuits, regardless of their specific architecture or depth. This technique has proven effective across diverse quantum applications, such as Grover's search, quantum amplitude estimation, the quantum support vector machine, and R\'enyi entropy measurements \cite{park2023qmp, rao2024quantum, baker2024qmp,  chowdhury2025probing, choi2026quantum, chowdhury2026quantum}.
Although QMP improves device utilization, concurrently executed circuits can interfere through crosstalk between neighbouring qubits, which is the principal drawback of co-resident execution~\cite{park2023qmp}. A common mitigation is to insert idle buffer qubits between co-resident circuits: Ohkura et al.~\cite{ohkura2022simultaneous} verified that even a single physical buffer qubit effectively suppresses crosstalk between adjacent circuits on IBM quantum computers. We take this isolation principle to its limit. Rather than packing the $K$ circuits densely with a one-qubit buffer between neighbours, we assign each to one of $K$ mutually routing-disconnected sets of $12$ physical qubits that occupy physically separated regions of the heavy-hex lattice and share no coupling path, separated by the device's remaining idle qubits, so that no entangling gate or routing SWAP can ever link two circuits. The physical-qubit layout of the four circuits is shown in Fig.~\ref{fig:qubit-layout} (``Qubit layout and reproducibility'' below).

A single MNIST sample evaluated through Eq.~\eqref{eq:pipeline} requires one circuit execution on the QPU.
At $75$-sample test-set scale with $256$ shots per circuit and $\mathcal{O}(10^1)$~s queue overhead per job, per-sample wall time is dominated by per-job overhead rather than quantum compute, and the $127$-qubit width of the device sits mostly idle.

QMP addresses this by packing $K$ identical copies of the trained circuit onto a wider QPU, executing $K$ samples in a single circuit invocation.
The QMP-packed circuit acts on
\begin{equation}
W \;=\; K\cdot n_q
\label{eq:qubits}
\end{equation}
physical qubits.
Inter-circuit isolation is provided not by buffer qubits but by \emph{physical separation on the heavy-hex topology}: the $K=4$ qubit layout uses four physically separated $12$-qubit circuits on \texttt{ibm\_yonsei} (Fig.~\ref{fig:qubit-layout}).
The $j$-th circuit ($j=1,\dots,K$) is encoded with the angle embedding of the $j$-th sample in the batch.
No entangling gate ever crosses circuit boundaries: the brick-wall CRX pattern is applied only within each $12$-qubit circuit, and inter-circuit gates would require SWAP routing that the transpiler does not introduce because the $K$ circuits are routing-disconnected in the circuit graph.

\subsubsection{Mathematical equivalence in the noiseless limit.}
Because no inter-circuit gate is ever applied, the joint state immediately before measurement factorizes:
\begin{equation}
|\Psi\rangle \;=\; |\psi_1\rangle \otimes |\psi_2\rangle \otimes \cdots \otimes |\psi_{K}\rangle,
\end{equation}
where $|\psi_j\rangle$ is the state that the $j$-th circuit would produce if executed alone on $n_q$ qubits with sample $j$'s angle embedding.
Single-qubit Pauli-$Z$ expectation values on any qubit belonging to the $j$-th circuit therefore reduce to expectations on $|\psi_j\rangle$ alone.
QMP is mathematically inert in the noiseless limit: the per-circuit predictions of the $K$-parallel circuit are bitwise identical to those of $K$ serially executed $n_q$-qubit circuits up to finite-shot noise.
We verified this on a noiseless simulator at $K\in\{1,2\}$: the maximum per-logit difference was $|\Delta|\le 2.6\times 10^{-6}$ across all $75$ test samples, with $100\%$ argmax agreement.
QMP is therefore expected to provide a throughput advantage on hardware (one job amortizes the per-job overhead across $K$ samples) but no accuracy advantage on classical simulators.
On a noisy QPU, any observed $K=1$ vs $K=K$ deviation must arise from finite-precision noise channels that scale with the number of active qubits and the depth of the compiled circuit --- specifically the larger physical-qubit footprint (4$\times$ more qubits at $K=4$, with potentially heterogeneous per-day error rates across the four circuits), the greater compiled circuit depth when the chosen circuits are not perfect linear chains on the heavy-hex topology, residual cross-circuit crosstalk under heavy simultaneous activity, and the seed-dependent assignment of test samples to circuits --- none of which are inter-circuit logical interactions in the circuit graph.

\subsubsection{Per-circuit expectation extraction.}
On hardware, per-circuit Pauli-$Z$ expectations are estimated from finite shot counts.
The \sampler{} primitive returns bitstring samples over all $W$ qubits per shot; the bitstrings are partitioned into $K$ circuit-specific slices using the qubit layout of Eq.~\eqref{eq:qubits}, and the Pauli-$Z$ expectation on the $i$-th qubit of the $j$-th circuit is estimated as
\begin{equation}
\widehat{\langle Z_{j,i}\rangle} \;=\; 1 \;-\; \frac{2\,n_{1,j,i}}{n_{\mathrm{shots}}},
\end{equation}
where $n_{1,j,i}$ is the count of shots in which the $i$-th bit of the $j$-th circuit's slice was measured to be $1$.

\subsubsection{Inference-only training.}
Training the QMP-packed model at $K\!>\!1$ on a classical simulator is infeasible because the $W$-qubit statevector grows as $2^W$.
We therefore separate training and inference: the trained checkpoint (Phase~1 + Phase~2) is produced at $K=1$ (single $12$-qubit circuit, $2^{12}=4096$-dimensional statevector), and the \emph{same checkpoint} is then evaluated at the deployment value of $K$ on the QPU.
Our hardware experiments use $(K,n_q)=(4,12)$, giving $W=48$ qubits occupying $\sim 38\%$ of the $127$-qubit device.

\section{Experiments}\label{sec:experiments}

All experiments in this section are carried out on \texttt{ibm\_yonsei}, a $127$-qubit IBM Quantum Eagle-r3 superconducting processor accessed through the IBM Qiskit Runtime cloud service.
In common with other Eagle-generation devices, it arranges its fixed-frequency transmon qubits on a heavy-hexagonal coupling lattice in which each qubit is connected to at most three neighbours, and realizes two-qubit entanglement through the native echoed cross-resonance (ECR) gate.
Its $127$-qubit width---far in excess of the $12$ qubits a single instance of our classifier requires---is precisely what makes the quantum-multi-programming inference experiments of Sec.~\ref{sec:qmp} possible.
The remainder of this section specifies the dataset, the five-ansatz model-selection pilot, the hardware deployment and error mitigation, the pinned physical-qubit layout and reproducibility controls, and the matched-capacity classical baseline, and then defines the five-cell experimental framework together with the evaluation protocol.

\subsection{Dataset and preprocessing}

Each experiment uses a deterministically-sampled subset of MNIST ($60{,}000$ training, $10{,}000$ test images): $N_{\mathrm{train}}=350$ training samples and $N_{\mathrm{val+test}}=150$ evaluation samples split $50/50$ into validation ($N_{\mathrm{val}}=75$) and test ($N_{\mathrm{test}}=75$).
Pixel values are converted to tensors scaled to $[0,1]$ and then standardized with the MNIST dataset mean $0.1307$ and standard deviation $0.3081$.
Mini-batches of $16$ samples are used everywhere.
Three independent draws of the data subset are produced by setting the global random seed to $2024$, $2025$, and $2026$, which simultaneously seeds the data sampler, the optimizer, the Aer shot-sampler, and the Qiskit transpiler routing (see ``Qubit layout and reproducibility'' below).
With a $75$-sample test set, the smallest distinguishable test-accuracy difference is one sample, equal to $1/75\approx 1.33\%$.

\subsection{Model-selection pilot}\label{sec:pilot}

Five candidate variational ans\"atze were evaluated under matched conditions (Phase~1 + Phase~2 on a noiseless statevector simulator, three seeds each) prior to the hardware experiments: \emph{VQC\_Fixed} (circular-ring CRX entanglement), \emph{VQC\_ZigZag} (brick-wall CRX entanglement, no wrap-around), \emph{VQC\_ZigZag\_Wrap} (brick-wall plus a wrap-around CRX per layer), and \emph{QCNN\_Hierarchical} at two scales ($8$~qubits with $2$~layers and $12$~qubits with $3$~layers; Cong-style convolution--pooling blocks with shared parameters).
Table~\ref{tab:pilot-metrics} summarizes the post-transpilation cost (compilation to the IBM basis $\{\mathrm{id},\rz,\sqrt{X},X,\mathrm{CX}\}$ under the heaviest available level of gate-merging and SWAP-routing optimization --- the level used for the hardware runs --- reporting gates-only depth excluding measurements and barriers, for the logical circuit with no device coupling map) and the resulting all-classical Phase~2 test accuracy.
Among the variational-quantum-classifier family, VQC\_ZigZag has the fewest CNOTs and the lowest depth; the hierarchical QCNN's convolution--pooling blocks act on a single qubit pair and are collapsed by aggressive two-qubit resynthesis to still fewer gates, but VQC\_ZigZag attains the highest mean test accuracy with perfect consistency across seeds, and is therefore selected as the deployed model (its architecture is detailed in Sec.~\ref{sec:zigzag-arch}).
The pilot evaluates expectation values analytically on a statevector simulator, whereas the five-cell framework (Sec.~\ref{sec:experiments}) uses the \sampler{} primitive at $256$ shots throughout, including for its noiseless reference (Cell~1). The two noiseless estimates of VQC\_ZigZag therefore differ in their measurement model rather than in the presence of noise, and their close agreement --- $78.7\pm 0.0\%$ analytically here versus $79.11\pm 2.04\%$ under finite-shot sampling in Table~\ref{tab:headline} --- indicates that $256$ shots suffice to reproduce the analytical decision boundary on this task, with the residual cross-seed spread attributable to shot noise.

\begin{table}[t]
\centering
\caption{Model-selection pilot. Post-transpilation metrics on the IBM basis $\{\mathrm{id},\rz,\sqrt{X},X,\mathrm{CX}\}$ under the heaviest available level of gate-merging and SWAP-routing optimization (gates-only depth, excluding measurements and barriers; logical circuit with no device coupling map, so routing SWAPs are not included). Test accuracy is the three-seed mean ($\pm$ standard deviation) on the $N_{\mathrm{test}}=75$ MNIST subset after Phase~1 + Phase~2 on a noiseless statevector simulator. Q-params is the number of trainable quantum parameters; C-params the trainable classical parameters in the encoder + readout.}
\label{tab:pilot-metrics}
\begin{tabular}{lccccc}
\toprule
Model & Depth & CNOTs & Q-params & C-params & Test ACC \\
\midrule
VQC\_Fixed       (12q, 3L) & 223 &  72 & 12 & 9{,}550 & $72.0\pm 2.7\%$ \\
\textbf{VQC\_ZigZag} (12q, 3L) & 45 & 66 & 15 & 9{,}550 & $\boldsymbol{78.7\pm 0.0\%}$ \\
VQC\_ZigZag\_Wrap (12q, 3L) &  45 &  72 & 18 & 9{,}550 & $77.8\pm 0.8\%$ \\
QCNN\_Hierarchical (8q,  2L) & 27 &  18 & 36 & 6{,}370 & $75.6\pm 2.0\%$ \\
QCNN\_Hierarchical (12q, 3L) & 38 & 30 & 54 & 9{,}550 & $77.8\pm 1.6\%$ \\
\bottomrule
\end{tabular}
\end{table}

\subsection{Quantum hardware deployment and error mitigation}

The measurement primitive is \sampler{}~\cite{ibm_qiskit_runtime} throughout: every Phase~2 training-time loss evaluation, every Hybrid Phase~2 hardware epoch, and every test-set forward pass uses the same primitive on \texttt{ibm\_yonsei}, which returns raw bitstring counts that are converted into per-circuit single-qubit Pauli-$Z$ expectation values via the partial-measurement procedure described above.

\subsubsection{Mitigation primitives.}
Two error-mitigation primitives, addressing distinct noise channels, are enabled in every hardware run:
\begin{itemize}
\item \emph{Dynamical decoupling} on idle qubits, which suppresses coherent dephasing accumulated during gate-execution gaps by inserting refocusing pulse sequences.
\item \emph{Pauli-twirled measurements}, which randomize the readout-error channel by inserting random Pauli operators immediately before each measurement~\cite{vandenberg2022trex}.
\end{itemize}
We chose \sampler{} as the framework-wide primitive precisely because QMP per-circuit extraction requires access to joint $W$-bit measurement outcomes; using the same Sampler-side mitigation across every cell of the framework eliminates any cross-primitive algorithmic-mitigation confound from the comparison.

\subsubsection{Transpilation and shot count.}
Compilation uses the heaviest available level of gate-merging and SWAP-routing optimization, with the transpiler's routing seed pinned to the experiment seed so that two runs of the same logical circuit on the same backend are compiled to the same physical gate sequence with the same SWAP placement.
Hardware and Aer-noiseless circuits are both executed with $256$ shots per circuit.

\subsection{Qubit layout and reproducibility}

\begin{figure}[t]\centering
\begin{subfigure}{0.49\textwidth}\centering
  \includegraphics[width=\textwidth]{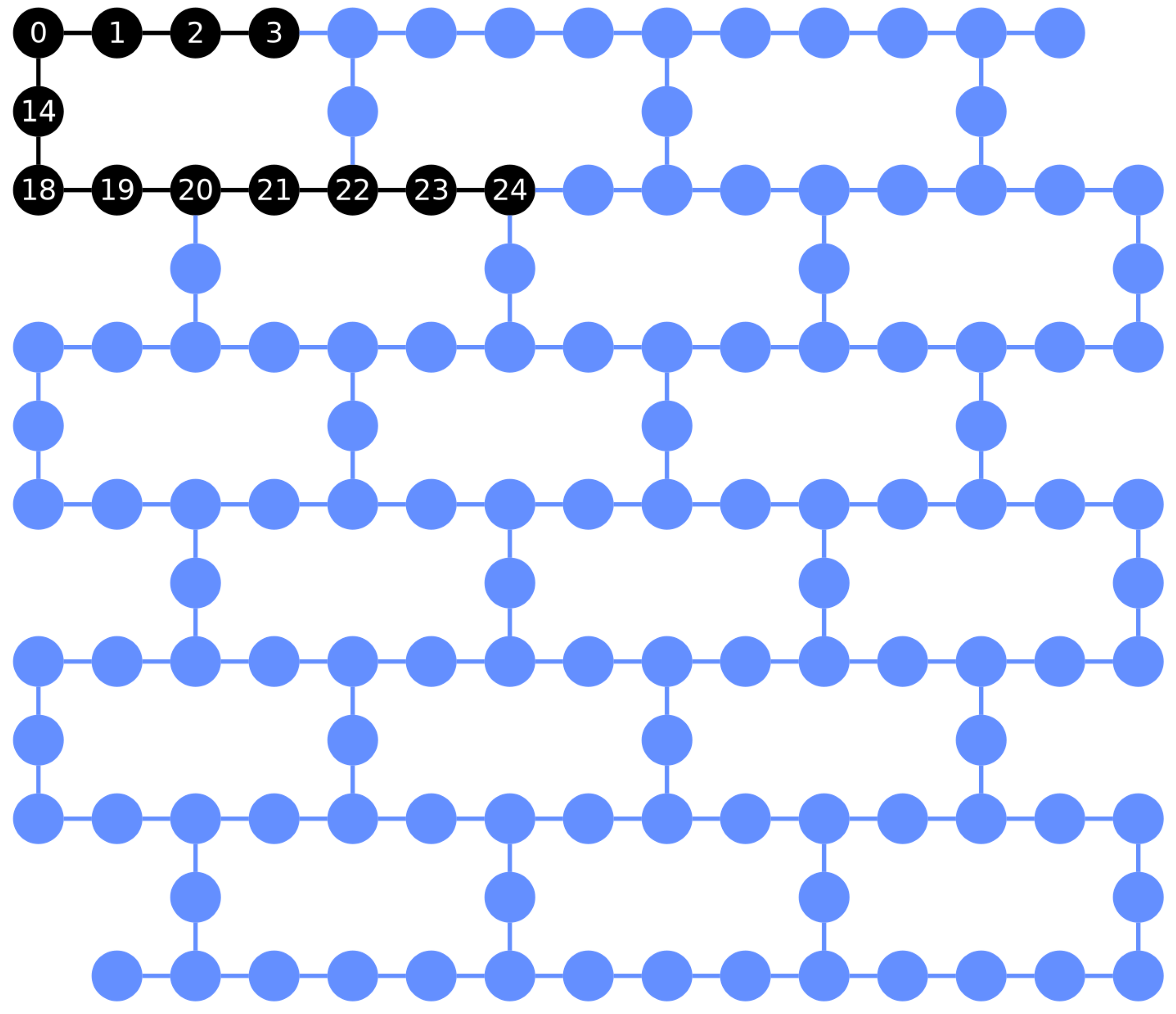}
  \caption{$K=1$: Circuit~1 (12 qubits).}\label{fig:layout-k1}\end{subfigure}\hfill
\begin{subfigure}{0.49\textwidth}\centering
  \includegraphics[width=\textwidth]{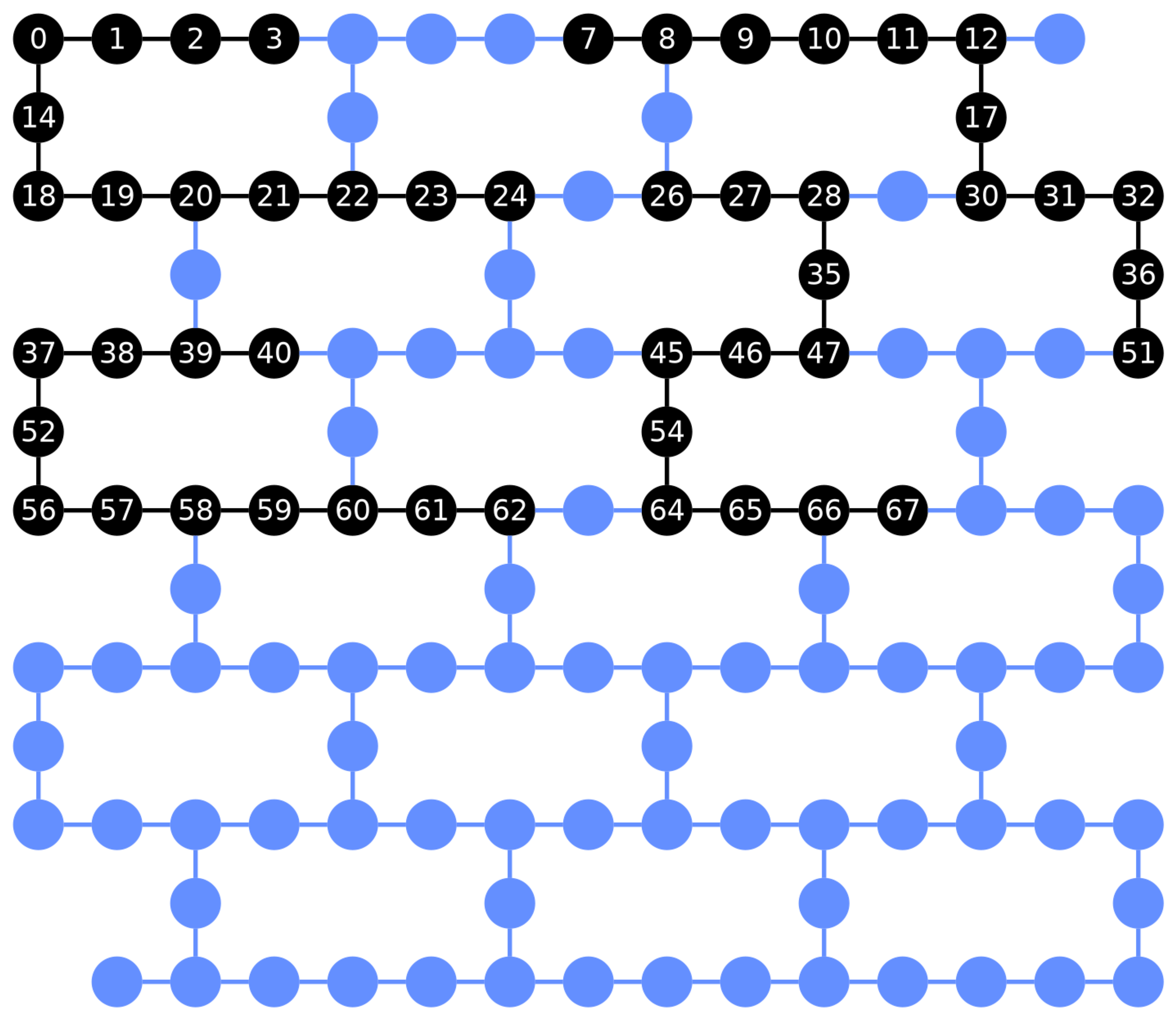}
  \caption{$K=4$ QMP: Circuits~1--4 (48 qubits).}\label{fig:layout-k4}\end{subfigure}
\caption{Pinned physical-qubit layout on the \texttt{ibm\_yonsei} heavy-hex topology (highlighted = active). (a) $K=1$ runs use Circuit~1; (b) $K=4$ QMP concatenates four routing-disconnected $12$-qubit circuits with no inter-circuit coupling. Circuit~1 (the $K=1$ layout) is reused as the first of the four circuits.}
\label{fig:qubit-layout}\end{figure}

\subsubsection{Pinned physical-qubit layout}
For every hardware run we pin the mapping from logical qubits to physical qubits on \texttt{ibm\_yonsei}, removing the run-to-run variability that would otherwise arise from the transpiler's daily-calibration-dependent qubit selection.
For $K=1$ runs the layout is a fixed set of $12$ physical qubits on the \texttt{ibm\_yonsei} heavy-hex topology; for $K=4$ QMP runs the layout is the concatenation of four such circuits, chosen so that the four circuits are routing-disconnected within the compiled circuit and no inter-circuit SWAP is ever introduced.
The chosen circuits are not perfect linear chains on the heavy-hex topology, so the transpiler introduces a small number of intra-circuit SWAPs within each $12$-qubit circuit; no SWAP ever crosses a circuit boundary.
The $K=1$ circuit is reused as the first circuit of the $K=4$ layout, which guarantees that any $K=1$-vs-$K=4$ contrast uses identical per-circuit physical-qubit placement.
The four circuits were selected in an error-aware manner from the device's reported calibration data rather than arbitrarily: every physical qubit in the layout satisfies a readout-assignment error below $1.044\times10^{-1}$, an identity-gate error below $1.505\times10^{-3}$, an $\sqrt{X}$ error below $1.505\times10^{-3}$, a Pauli-$X$ error below $1.505\times10^{-3}$, and a measurement error below $1.044\times10^{-1}$, while every two-qubit connection used satisfies an ECR error below $2.774\times10^{-2}$.
This keeps the per-circuit noise profile comparable across the four QMP circuits, so that any $K=1$-vs-$K=4$ accuracy difference reflects the aggregate cost of activating four times as many low-error qubits rather than the inclusion of an outlier-noisy qubit or link.
Both layouts are shown in Fig.~\ref{fig:qubit-layout}.

\subsubsection{Reproducibility seeding}
Every run pins the transpiler's SWAP-routing seed, the Aer simulator's shot-sampling state, and the COBYLA optimizer's internal random state to a single experiment seed (alongside the standard global random-number generator seeding for the data sampler and PyTorch).
Together with the pinned physical-qubit layout above, this makes every noiseless run bitwise reproducible and every hardware run reproducible up to (i) finite-shot Monte Carlo noise and (ii) day-to-day calibration drift on the physical device.
We mitigate calibration drift by submitting all $15$ framework runs as a tight temporal cluster (target: same day; minimum: same week).

\subsection{Matched-capacity classical baseline}

To assess whether the quantum module provides any per-parameter advantage at this scale, we benchmark the trained quantum classifier against a matched-capacity classical multilayer perceptron (MLP).
The classical MLP has the architecture $\textsf{Linear}(784\!\to\!12)\,+\,\mathrm{ReLU}\,+\,\textsf{Linear}(12\!\to\!10)$, which preserves the $784\!\to\!12$ encoder and $12\!\to\!10$ readout of the quantum pipeline and replaces the $12$-qubit variational quantum circuit with a single ReLU non-linearity.
The total parameter count is $9{,}550$, matching the quantum pipeline's combined ($\boldsymbol{\phi}+\boldsymbol{\theta}$) count of $9{,}565$ to within $0.16\%$.
Training is via Adam (learning rate $10^{-3}$, weight decay $10^{-4}$) for $50$ epochs on the same MNIST $350/75/75$ split at each of the three seeds.

\subsection{Experimental framework: five hardware-deployment cells}

The headline experimental design is a five-cell table that holds the model, data split, optimizer, batch size, measurement primitive, shot count, mitigation, qubit layout, and all four reproducibility seeds constant across cells.
The cells differ only in (i) the device used during Phase~2 training and (ii) the device used during the test evaluation.
This design ensures that every cell-to-cell contrast isolates exactly one effect.

\begin{table}[t]
\centering
\caption{The five hardware-deployment cells. Phase~1 (Adam on a noiseless statevector simulator, classical layers only) is identical across all cells. Phase~2 device and test device vary as shown. Cells 1--3 share one Aer-noiseless Phase~2 checkpoint per seed; Cells 4--5 share one Hybrid Phase~2 checkpoint per seed. The same trained weights are reused across cells in the same group, so no Phase~2 training is repeated.}
\label{tab:cells}
\begin{tabular}{cll}
\toprule
Cell & Phase~2 training device & Test device \\
\midrule
1 & Aer-noiseless (\sampler{}, $256$ shots) & Aer-noiseless, $K=1$ \\
2 & Aer-noiseless (\sampler{}, $256$ shots) & \texttt{ibm\_yonsei}, $K=1$ ($12$ qubits) \\
3 & Aer-noiseless (\sampler{}, $256$ shots) & \texttt{ibm\_yonsei}, $K=4$ QMP ($48$ qubits) \\
4 & Hybrid: $18$ Aer-noiseless + $2$ \texttt{ibm\_yonsei}    & \texttt{ibm\_yonsei}, $K=1$ ($12$ qubits) \\
5 & Hybrid: $18$ Aer-noiseless + $2$ \texttt{ibm\_yonsei}    & \texttt{ibm\_yonsei}, $K=4$ QMP ($48$ qubits) \\
\bottomrule
\end{tabular}
\end{table}

\subsubsection{Contrasts.}
The pairwise comparisons that the framework supports are:
\begin{itemize}
\item Cell~$1\!\leftrightarrow\!2$ isolates hardware noise on a single forward pass at test, with the same trained weights and at $K=1$.
\item Cell~$2\!\leftrightarrow\!3$ isolates the $K=4$ QMP packing penalty on hardware ($48$ qubits vs.\ $12$), with identical mitigation and trained weights.
\item Cell~$2\!\leftrightarrow\!4$ isolates the effect of two hardware-training COBYLA epochs on $K=1$ test accuracy.
\item Cell~$3\!\leftrightarrow\!5$ isolates the same hardware-training effect at $K=4$.
\item Cell~$4\!\leftrightarrow\!5$ isolates the $K=4$ QMP packing penalty for hybrid-trained weights.
\end{itemize}

\subsubsection{Replication and reuse.}
Each cell is run at three independent seeds ($2024,2025,2026$), giving $15$ total measurements.
Critically, cells in the same Phase~2 training group reuse one trained checkpoint per seed: Cells $1$--$3$ share the Aer-noiseless Phase~2 checkpoint, and Cells $4$--$5$ share the Hybrid Phase~2 checkpoint.
The shared-checkpoint design guarantees that within-group cell-to-cell contrasts compare \emph{bitwise-identical trained weights} evaluated under different test-time devices, so the resulting accuracy differences are uniquely attributable to the device variable.
Per seed, this is $2$ training runs and $3$ inference runs; across $3$ seeds, this is $6$ training runs and $9$ inference runs ($15$ jobs total), with zero redundant Phase~2 retraining.

\subsection{Evaluation protocol}

For every (cell, seed) combination, we report classification accuracy on the $N_{\mathrm{test}}=75$-sample held-out test set together with the per-batch wall time, per-sample logits, predictions, and ground-truth labels.
We additionally compute the matched-capacity classical MLP baseline for a direct per-parameter capacity comparison.

\section{Results}\label{sec:results}

\subsection{Overview of the measurements}

\begin{figure}[t]\centering
\includegraphics[width=\textwidth]{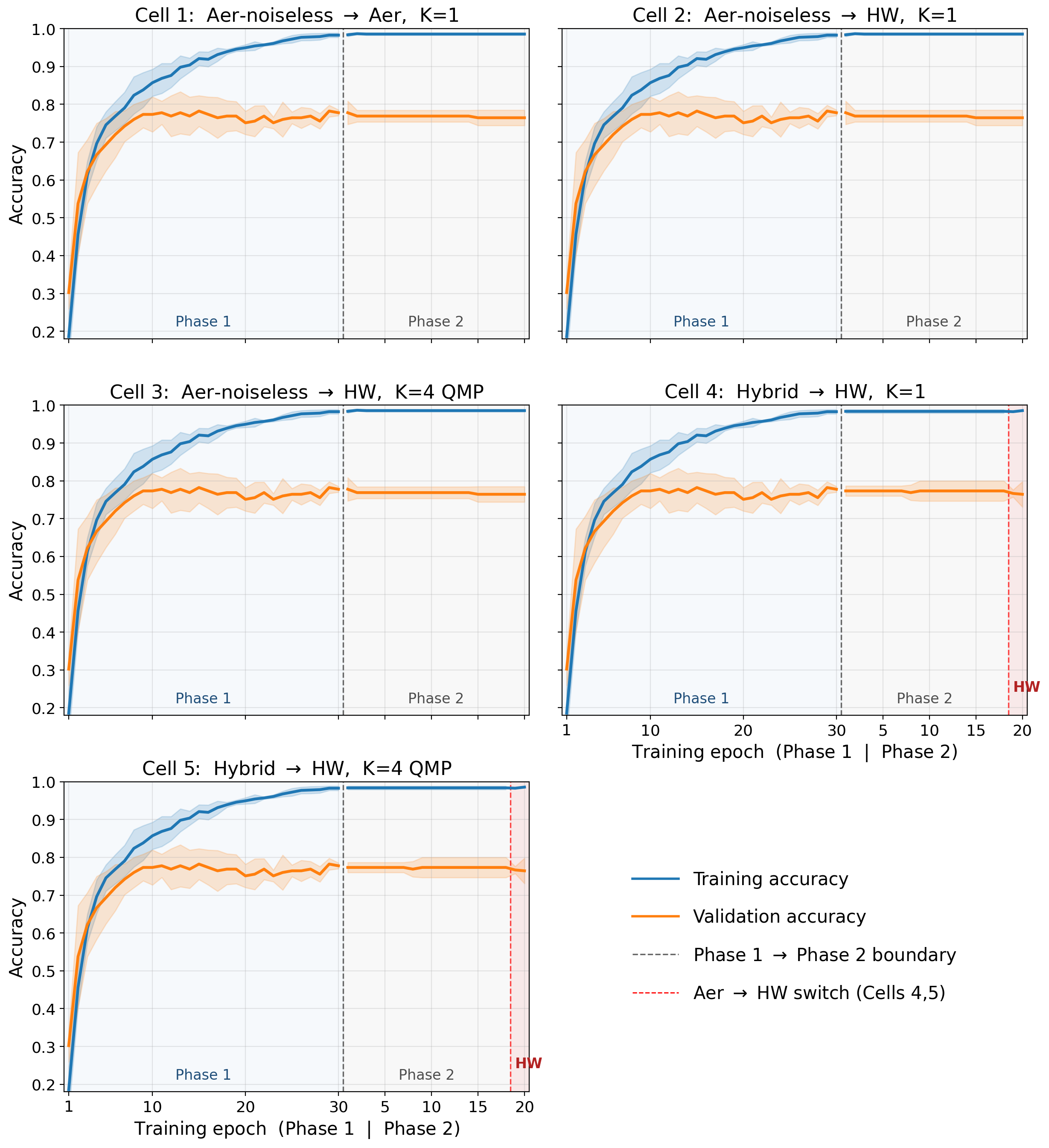}
\caption{Two-phase training curves for VQC\_ZigZag ($12$q/$3$L, MNIST $350/75/75$; mean $\pm$ s.d.\ over three seeds). Phase~1 (Adam, $30$ epochs, classical layers) raises validation accuracy to $\sim\!78\%$, while Phase~2 (COBYLA, $20$ epochs, quantum parameters) is essentially flat in every cell. The red dashed line marks the Aer$\,\rightarrow\,$hardware switch at Phase~2 epoch~$19$ for the hybrid-trained Cells~4--5. The flat Phase~2 traces visualize the central result of Claim~3: gradient-free quantum-parameter optimization, whether on a simulator or on hardware, does not move the classification decisions.}
\label{fig:learning-curves}\end{figure}

We deployed the selected VQC\_ZigZag classifier end-to-end on the IBM Eagle-r3 device \texttt{ibm\_yonsei} across the five-cell experimental framework described in Methods.
This yields $15$ test-set measurements, complemented by a noiseless-simulator reference (Cell~1) and a matched-capacity classical MLP baseline trained and evaluated under the identical data protocol.
All accuracies are reported on the held-out $N_{\mathrm{test}}=75$-sample MNIST test set, and every summary quantity below --- accuracies and wall times alike --- is the three-seed mean $\pm$ standard deviation.
Because one test sample equals $1/75\approx 1.33\%$, the smallest distinguishable accuracy difference is one sample; throughout, we treat differences below roughly two samples ($\approx 2.7$ percentage points, pts) as \emph{within sampling noise}.
The five-cell design holds the model, data split, optimizer, measurement primitive (\sampler{}, $256$ shots), error mitigation, qubit layout, and all reproducibility seeds constant; cells differ only in the Phase~2 training device and the test device.
Critically, the cells within each training group share \emph{bit-identical} trained weights (Cells~1--3 share one simulator-trained checkpoint per seed; Cells~4--5 share one hybrid-trained checkpoint per seed), so every within-group cell-to-cell contrast isolates exactly one variable.
Table~\ref{tab:headline} reports the headline accuracies and Table~\ref{tab:contrasts} the isolated contrasts; Fig.~\ref{fig:learning-curves} shows the underlying two-phase training dynamics for all five cells.

\begin{table}[t]
\centering
\caption{Headline test accuracy ($N_{\mathrm{test}}=75$ MNIST; three-seed mean $\pm$ standard deviation) for the five hardware-deployment cells, with the matched-capacity classical MLP baseline. Cells~1--3 are evaluated from one shared simulator-trained (Aer-noiseless) checkpoint per seed; Cells~4--5 from one shared hybrid-trained ($18$ Aer-noiseless $+\,2$ \texttt{ibm\_yonsei} COBYLA epochs) checkpoint per seed. Phase~1 is identical across all cells.}
\label{tab:headline}
\begin{tabular}{clc}
\toprule
Cell & Train $\rightarrow$ Test & Mean $\pm$ Std \\
\midrule
1 & Aer-noiseless $\rightarrow$ Aer-noiseless, $K=1$        & $79.11 \pm 2.04$ \\
2 & Aer-noiseless $\rightarrow$ HW, $K=1$                    & $73.78 \pm 1.54$ \\
3 & Aer-noiseless $\rightarrow$ HW, $K=4$ QMP                & $76.44 \pm 4.68$ \\
4 & Hybrid $\rightarrow$ HW, $K=1$                           & $76.89 \pm 0.77$ \\
5 & Hybrid $\rightarrow$ HW, $K=4$ QMP                       & $74.67 \pm 3.53$ \\
\midrule
\multicolumn{2}{l}{Classical MLP ($9{,}550$ params)}            & $82.22 \pm 3.36$ \\
\bottomrule
\end{tabular}

\vspace{2pt}
{\footnotesize All values in \%. ``HW'' denotes execution on \texttt{ibm\_yonsei}.}
\end{table}

\begin{table}[t]
\centering
\caption{Isolated cell-to-cell contrasts (three-seed mean of cell means). Each contrast holds every variable but one fixed and, within a training group, compares bit-identical trained weights. ``$\approx$ samples'' converts the gap to $75$-sample test-set units ($1$ sample $=1.33$ pts). Differences below $\approx 2$ samples are within sampling noise.}
\label{tab:contrasts}
\begin{tabular}{llccl}
\toprule
Contrast & Variable isolated & $\Delta$ (pts) & $\approx$ samples & Reading \\
\midrule
Cell 1 $\rightarrow$ 2 & Hardware noise, $K=1$               & $-5.33$ & $\sim 4$ & meaningful \\
Cell 1 $\rightarrow$ 3 & Hardware noise $+$ $K=4$ (combined) & $-2.67$ & $\sim 2$ & within noise \\
Cell 2 $\rightarrow$ 3 & $K=4$ QMP packing (sim weights)     & $+2.67$ & $\sim 2$ & within noise \\
Cell 4 $\rightarrow$ 5 & $K=4$ QMP packing (hybrid weights)  & $-2.22$ & $\sim 2$ & within noise \\
Cell 2 $\rightarrow$ 4 & On-hardware training benefit, $K=1$ & $+3.11$ & $\sim 2$ & within noise \\
Cell 3 $\rightarrow$ 5 & On-hardware training benefit, $K=4$ & $-1.78$ & $\sim 1$ & within noise \\
\bottomrule
\end{tabular}
\end{table}

\subsection{A 12-qubit QML model classifies MNIST end-to-end on real hardware}

The trained $12$-qubit VQC executes end-to-end on \texttt{ibm\_yonsei} and produces usable classification.
On the noiseless reference (Cell~1) the model reaches $79.11\pm 2.04\%$; deploying the \emph{same trained weights} to hardware for a single forward pass at $K=1$ (Cell~2) gives $73.78\pm 1.54\%$.
Because Cells~1 and~2 share bit-identical weights and differ only in the test device, the Cell~1$\,\rightarrow\,$2 gap of $-5.33$ pts ($\sim 4$ test samples; Table~\ref{tab:contrasts}) is a clean estimate of the single-forward-pass hardware-noise penalty under dynamical decoupling and Pauli-twirled-measurement mitigation.
The hardware result is also reproducible: the cross-seed standard deviation at $K=1$ on hardware is only $1.54\%$ (about one sample), among the tightest in Table~\ref{tab:headline}.
We conclude that a non-trivial $12$-qubit quantum classifier runs on a current superconducting device with a modest and stable accuracy cost, establishing feasibility.

\subsection{Quantum Multi-Programming delivers $K$-parallel inference at no mean-accuracy cost}

We packed $K=4$ identical copies of the trained circuit onto $48$ physically separated qubits of \texttt{ibm\_yonsei} and read out four samples per circuit invocation --- to our knowledge the first application of Quantum Multi-Programming~\cite{park2023qmp,baker2024qmp} to a trained QML model.
Because no entangling gate crosses a circuit boundary, the $K$-circuit packing is mathematically equivalent to $K$ independent single circuits in the noiseless limit (verified in Methods to a per-logit difference $\le 2.6\times 10^{-6}$ with $100\%$ argmax agreement); any $K=1$ vs $K=4$ difference observed on hardware therefore arises from finite-precision noise channels, not from logical interaction.

On the shared simulator-trained weights, moving from $K=1$ (Cell~2, $73.78\%$) to $K=4$ QMP (Cell~3, $76.44\%$) changes the mean by $+2.67$ pts ($\sim 2$ samples) --- within sampling noise (Table~\ref{tab:contrasts}).
The correct characterization of the $K=4$ packing effect is therefore \emph{variance amplification, not mean degradation}: the mean is statistically unchanged while the cross-seed standard deviation grows about three-fold, from $1.54\%$ at $K=1$ to $4.68\%$ at $K=4$ (Table~\ref{tab:variance}).
The same pattern holds for the hybrid-trained weights (Cell~4 $\rightarrow$ 5): the mean changes by only $-2.22$ pts ($\sim 2$ samples, within noise) while the cross-seed standard deviation grows from $0.77\%$ at $K=1$ to $3.53\%$ at $K=4$.
The added variance is consistent with $K=4$ activating four times as many physical qubits --- three of the four circuits run with their own per-day calibration profiles, and the seed-dependent assignment of samples to circuits adds a (sample\,$\times$\,circuit) variance component absent at $K=1$ --- and is mitigable by replication and averaging.

The practical payoff is throughput, and the framework measures it directly. Cells~2 and~3 evaluate bit-identical trained weights on the same $75$-sample test set and differ only in $K$, so their end-to-end wall times isolate the effect of packing. Evaluating the full test set takes $5225$\,s at $K=1$ and $1388$\,s at $K=4$, an end-to-end speedup of $3.76\times$ ($3.69$--$3.82$) against the theoretical ceiling of $K=4$. Excluding the first batch of each run, which additionally carries one-time session-establishment and compilation overhead, the steady-state per-batch speedup is $3.96\times$ ($3.94$--$3.98$), i.e.\ essentially the full $K$-fold gain. The two cells were submitted as separate jobs rather than interleaved, so device load was not held fixed across the comparison; the $0.03\times$ spread is far tighter than queue variation would produce. QMP therefore converts $K$ samples' worth of per-job overhead into one and realizes close to the full $K$-fold throughput gain, at a mean-accuracy cost that is statistically zero on this task and scale.

\begin{table}[t]
\centering
\caption{Variance amplification under $K=4$ QMP packing. The cross-seed mean is essentially unchanged within sampling noise while the standard deviation grows by roughly $3$--$5\times$.}
\label{tab:variance}
\begin{tabular}{lcccc}
\toprule
Training protocol & \multicolumn{2}{c}{Mean test ACC (\%)} & \multicolumn{2}{c}{Std across seeds (\%)} \\
\cmidrule(lr){2-3}\cmidrule(lr){4-5}
 & $K=1$ & $K=4$ & $K=1$ & $K=4$ \\
\midrule
Simulator-trained (Cells 2, 3) & $73.78$ & $76.44$ & $1.54$ & $4.68$ \\
Hybrid-trained (Cells 4, 5)    & $76.89$ & $74.67$ & $0.77$ & $3.53$ \\
\bottomrule
\end{tabular}
\end{table}

\subsection{On-hardware optimization adds no accuracy: train on a simulator, deploy for inference only}

To test whether spending scarce QPU time on in-the-loop optimization pays off, we compare hybrid training (two of twenty COBYLA epochs executed on \texttt{ibm\_yonsei}; Cells~4--5) against simulator-only training (Cells~2--3), at matched test device and matched $K$.
The two hardware-training epochs produce no measurable change in test accuracy: $+3.11$ pts at $K=1$ (Cell~2$\,\rightarrow\,$4, $\sim 2$ samples) and $-1.78$ pts at $K=4$ (Cell~3$\,\rightarrow\,$5, $\sim 1$ sample), both within sampling noise (Table~\ref{tab:contrasts}).
The cross-seed standard deviations are comparable between the two training protocols ($0.77\%$ vs $1.54\%$ at $K=1$; $3.53\%$ vs $4.68\%$ at $K=4$), so the two hardware epochs change neither the mean accuracy nor the run-to-run stability in a detectable way.
This is consistent with the flat Phase~2 optimization visible in Fig.~\ref{fig:learning-curves}: the COBYLA loss computed from noisy, shot-based expectation values drifts without ever moving the classification decisions, whether the optimizer runs on a simulator or on real hardware.

The conclusion is unambiguous: there is no payoff to in-the-loop QPU optimization, so the practical NISQ workflow is to \textbf{train the quantum model on a classical simulator and reserve the hardware for inference} --- the stage that QMP accelerates. The relevant asymmetry is not one of aggregate lifetime cost but of noise sensitivity, and it follows from what each stage demands of the same noisy measurement. Optimization must resolve the difference in loss between nearby parameter settings, a quantity that shrinks as the search converges; inference need only preserve the argmax over ten logits, which tolerates perturbations of any magnitude that leave the ranking intact. The same shot and device noise therefore degrades the training signal well before it degrades the prediction --- exactly the pattern observed here, where Phase~2 optimization moves the loss without changing a single classification decision (Fig.~\ref{fig:learning-curves}) while a full forward pass on hardware costs only $5.33$~pts.

\subsection{Matched-capacity classical baseline}

To situate these results, we benchmark the quantum classifier against a classical MLP matched to the same parameter budget ($9{,}550$ trainable parameters vs.\ the quantum pipeline's $9{,}565$; Methods).
The classical baseline reaches $82.22\pm 3.36\%$ (Table~\ref{tab:headline}), exceeding the best quantum cell (Cell~1, noiseless, $79.11\%$) by $\sim 3$ pts and the best hardware cell (Cell~4, $76.89\%$) by $\sim 5$ pts.
At this scale, therefore, the variational quantum module confers no per-parameter accuracy advantage over a classical network of equal capacity.
We accordingly frame the contribution of this work not as a quantum accuracy advantage but as a feasibility-and-workflow result: a non-trivial trained QML model can be executed, parallelized via QMP, and deployed for inference on current quantum hardware, with the empirical guidance that classical-simulator training followed by hardware inference is the workflow that current devices reward.

\section{Conclusion and Perspectives}\label{sec:conclusion}
We have presented a unified, hardware-deployed framework for multi-class quantum image classification and used it to run ten-class MNIST end-to-end on a $127$-qubit IBM Eagle-r3 processor (\texttt{ibm\_yonsei}).
Within a single tightly controlled protocol---a fixed model, data split, optimizer, measurement primitive, mitigation pipeline, qubit layout, and reproducibility seeds, with cells differing only in the training and test device---the framework jointly addresses the four obstacles that have largely confined quantum machine learning to noiseless simulation and binary tasks: classification scale, training cost, inference throughput, and ansatz selection.

A controlled comparison of five variational ans\"atze, pairing classification accuracy with post-transpilation circuit depth and entangling-gate count on the native IBM gate set, identified a brick-wall-entanglement circuit (VQC\_ZigZag; $12$ qubits, $15$ trainable quantum parameters) as the best accuracy--cost trade-off.
Deployed on hardware, this classifier runs end-to-end and produces usable ten-class predictions: relative to its noiseless reference ($79.11\%$), a single forward pass on the QPU at $K=1$ incurs a modest and highly reproducible hardware-noise penalty of $5.33$ percentage points (cross-seed standard deviation $1.54\%$, about one test sample).
This establishes that a non-trivial $12$-qubit quantum classifier can be executed on a current superconducting device at a stable, quantifiable accuracy cost.

To our knowledge, this is also the first application of QMP to a trained QML model.
Packing $K=4$ identical copies of the trained circuit onto $48$ physically separated qubits, we read four samples per circuit invocation and reduce the number of QPU job submissions fourfold.
Because no entangling gate crosses a circuit boundary, the packed circuit is mathematically equivalent to serial execution in the noiseless limit (verified to a per-logit difference $\le 2.6\times10^{-6}$ with $100\%$ argmax agreement); on hardware, $K=4$ packing leaves the mean test accuracy statistically unchanged---within sampling noise---and instead amplifies cross-seed variance roughly three- to fivefold.
The correct characterization is therefore \emph{variance amplification, not mean degradation}.
Since per-sample wall time on cloud-accessed hardware is dominated by per-job queue and submission overhead rather than quantum computation, QMP converts $K$ samples' worth of overhead into one at no mean-accuracy cost---a structural throughput gain for the inference stage that is mitigable, where variance matters, by replication and averaging.

Our two-phase protocol separates a gradient-based classical optimization of the encoder and readout from a gradient-free COBYLA optimization of the quantum parameters, sidestepping both the parameter-shift cost of on-hardware gradients and the barren-plateau pathology of joint optimization~\cite{mcclean2018barren}.
A controlled comparison then yielded a clear negative result: two epochs of on-hardware COBYLA fine-tuning produced no measurable change in test accuracy (within sampling noise at both $K=1$ and $K=4$) and no detectable change in cross-seed stability, consistent with a COBYLA loss computed from noisy shot-based expectation values that drifts without ever moving the classification decisions.
Synthesizing the feasibility, QMP, and training results, we advocate a concrete NISQ workflow---\textbf{train the quantum model on a classical simulator, then reserve the hardware exclusively for inference}---the stage that QMP accelerates and, under inference-only deployment, the only stage that need engage the quantum device at all.

We deliberately calibrated these results against a matched-capacity classical multilayer perceptron, which reached $82.22\%$ and exceeded the best quantum cell (the noiseless reference, $79.11\%$) by roughly three percentage points.
At this scale, therefore, the variational quantum module confers no per-parameter accuracy advantage, and we frame our contribution accordingly: a feasibility-and-workflow demonstration rather than a claim of quantum advantage.
The study is bounded by its scale---a $350/75/75$ MNIST subset whose $75$-sample test set resolves accuracy only to $\approx 1.33\%$ per sample, three random seeds, and a single device over a limited calibration window.
Closing the gap to classical performance, and determining whether quantum feature maps can offer an advantage that grows with problem size, will require larger datasets and qubit counts, deeper error mitigation or partial error correction, and QMP at higher packing factors.
For each of these directions, the controlled, shared-weight experimental template introduced here---which isolates hardware noise, on-hardware training, and packing effects one variable at a time---provides a reusable and honest basis for measurement as hardware advances toward the fault-tolerant regime~\cite{preskill2018quantum}.

\section*{Acknowledgment}
\noindent
K.Y. is supported by the Brookhaven National Laboratory LDRD 24-061, 25-033, and the U.S. Department of Energy, Office of Science, Grants No. DE-SC0012704.
This research was supported by the Institute for Convergence Research and Education in Advanced Technology (iCREATE) and the Institute of Quantum Information Technology (IQIT) at Yonsei University through the use of the Yonsei quantum computing system, by the APEC-APRU partnership program (B0080429002759) funded by the Ministry of Education, Korea, by the National Research Foundation of Korea (NRF) grant funded by the Korea government (MSIT) (No. 2021R1C1C1006503, RS-2023-00266787, RS-2023-00265406, RS-2024-00421268, RS-2024-00342301, RS-2024-00435727, RS-2025-25457239, RS-2021-NR061370, NRF-2021M3E5D2A01022515, NRF-2021S1A3A2A02090597, and RS-2025-00514606), and by the Researchers Program through Seoul National University (No. 200-20250071, 200-20250049, 200-20250116, 200-20250115, 200-20250113, 0670-20250039, 200-20260009). Additional support was provided by the Institute of Information \& Communications Technology Planning \& Evaluation (IITP) grant funded by the Korea government (MSIT) [No. RS-2021-II211343, Artificial Intelligence Graduate School Program, Seoul National University] and by the Global Research Support Program in the Digital Field (RS-2024-00421268). This work was also supported by the Artificial Intelligence Industrial Convergence Cluster Development Project funded by the Ministry of Science and ICT and Gwangju Metropolitan City, by the Korea Brain Research Institute (KBRI) basic research program (25-BR-05-01), by the Korea Health Industry Development Institute (KHIDI) and the Ministry of Health and Welfare, Republic of Korea (HR22C1605), and by the Korea Basic Science Institute (National Research Facilities and Equipment Center) grant funded by the Ministry of Education (RS-2024-00435727). We acknowledge the National Supercomputing Center for providing supercomputing resources and technical support (KSC-2023-CRE-0568, KSC-2024-CRE-0198, KSC-2025-CRE-0340).
We acknowledge the Yonsei University Quantum Computing Project Group for providing support and access to the Quantum System One (Eagle Processor), which is operated at Yonsei University.
An award for computer time was provided by the U.S. Department of Energy’s (DOE) ASCR Leadership Computing Challenge (ALCC). This research used resources of the National Energy Research Scientific Computing Center (NERSC), a DOE Office of Science User Facility under Contract No. DE-AC02-05CH11231 using NERSC award DDR-ERCAP0037323, under ALCC award m4750-2024, and supporting resources at the Argonne and Oak Ridge Leadership Computing Facilities, U.S. DOE Office of Science user facilities at Argonne National Laboratory and Oak Ridge National Laboratory.

\clearpage

\bibliography{references}

\end{document}